\documentclass[runningheads]{llncs}
\usepackage{comment}
\usepackage{graphicx}
\graphicspath{{./}}
\usepackage{multirow}
\usepackage{booktabs}
\usepackage{amsmath}
\usepackage{amsfonts}
\usepackage{adjustbox}
\usepackage{placeins}
\usepackage{subcaption}
\begin{document}
\title{MSTF-Net: A UAV-Oriented Multi-Spectral Video Segmentation Method via Modality-Robust, Scale-Adaptive, and Consistent Fusion}

\def\CICAISubNumber{100}  
\titlerunning{MSTF-Net}
\authorrunning{C. Wang et al.}
\author{Chenwei Wang\thanks{These authors contributed equally to this work. Corresponding author: Zelin Li (chiellini.lee@gmail.com).} \and
Zhida Wang\textsuperscript{$\star$} \and
Zelin Li\textsuperscript{$\star$} \and
Elif Ozden-Yenigun \and
Houde Liu}
\institute{}

\maketitle

\begin{abstract}
Multi-spectral video segmentation is essential for robust scene understanding in unmanned aerial vehicle (UAV) applications such as city planning, land use monitoring, traffic monitoring, and crowd estimation. While the fusion of RGB and thermal modalities offers complementary information for perception under varying lighting and visibility conditions, two fundamental challenges remain: (1) the modal fusion dilemma, arising from significant discrepancies between RGB and thermal features that obscure complementary cues, and (2) temporal variation, induced by rapid motion and viewpoint changes on UAV platforms, which leads to appearance inconsistency and misalignment across frames.
To address these issues, this study proposed MSTF-Net, a modality-robust scale-adaptive fusion framework for multi-spectral video segmentation that effectively models cross-modal fusion and temporal consistency. The Modality Spatial Complementary Suppression and Enhancement (MSCSE) module generates unified instance queries via cross-modal attention and suppresses modality-specific noise using residual-guided discrepancy filtering and consistency constraints. To model temporal dynamics, the Multi-scale Temporal Cross-modality Semantic Consistency (MTCSC) module adaptively adjusts the temporal receptive field based on frame distance, capturing both coarse global context and fine local structure across time.
Extensive ablation experiments on public RGB-T datasets demonstrate that MSTF-Net achieves state-of-the-art segmentation performance, especially under challenging conditions such as small targets, occlusion, and modality degradation. Specifically, reached 56.42\% mIoU on the MVSeg dataset and 51.80\% mIoU on the CART dataset, respectively.

\end{abstract}
\section{Introduction}
\label{sec1}

The rapid advancement of UAVs has made them vital for tasks ranging from environmental monitoring to disaster relief \cite{ezequiel2014uav}. Their mobility enables access to remote or hazardous areas, such as earthquake zones, and supports everyday uses like infrastructure surveillance and urban deliveries \cite{mohsan2023unmanned}, along with related aerial and device-free sensing tasks such as UAV localization, activity inspection, and radar-based urban perception \cite{yin2026ciuav,yin2025spatio,wang2019parking}. However, their heavy reliance on visual cameras limits performance in adverse conditions (e.g., fog, darkness, overexposure), posing a major obstacle to their deployment in many safety-critical missions \cite{mohsan2022towards} like low-altitude rescue. 
To expand the capabilities of low-altitude UAVs, multi-spectral video semantic segmentation (MVSS) \cite{dong2022mvss} has emerged as a promising approach. By fusing RGB and thermal data, MVSS leverages the strengths of both RGB and thermal imaging. 

\begin{figure}[htbp]
  \centering
  \includegraphics[width=0.95\linewidth]{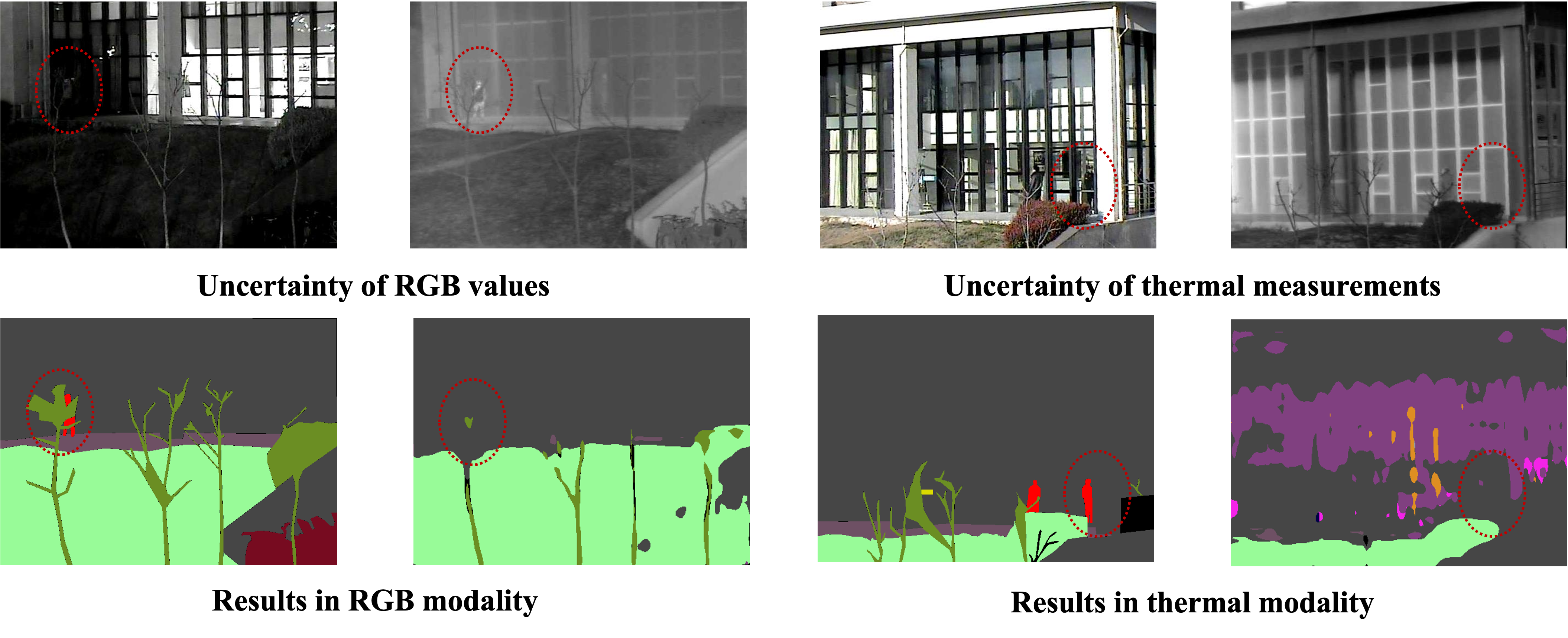}
  \caption{The challenge of UAV-based multi-spectral video segmentation: For RGB (left) versus thermal (right), input image and corresponding uncertainty map with red dashed circles highlight regions of high uncertainty. Semantic segmentation outputs with the same circled areas show typical misclassifications. RGB misses fine structures, while thermal struggles with complex backgrounds.}
  \label{FIRST_IMP}
\end{figure}

However, multi-spectral video fusion not only suffers from the multi-modal fusion challenge \cite{tang2023comparative}, but also faces additional challenges \cite{li2022multi} introduced by the dynamic and fast-moving UAV platform, as shown in Fig. \ref{FIRST_IMP}.
At the modality level, effectively fusing RGB and thermal data requires the extraction of truly complementary information (TCI) from each modality \cite{fei2023uav}. RGB images provide high-level semantic features such as texture and color, while thermal imagery captures heat distribution independent of lighting conditions \cite{maimaitijiang2020crop}. Due to their inherently different feature spaces which lead a phenomenon known as heterogeneous feature representation \cite{dong2020heterogeneous}, the two modalities often contain significant differences that are not necessarily complementary. These differences can introduce fusion noise \cite{sun2022mlr} rather than TCI. Thus, identifying and preserving TCI while suppressing irrelevant or conflicting signals in both spatial and temporal dimensions remains a major challenge in current research \cite{zhang2021image}.
At the platform level, the high mobility of UAVs results \cite{cheng2024methods} in frequent changes in altitude and camera orientation. These dynamic variations lead to rapid shifts in object size, occlusions, and scene semantics, complicating accurate segmentation. Small objects may be easily missed \cite{liu2020uav}, while larger or more complex targets often suffer from poor boundary definition or structural distortion \cite{morgenthal2014quality} due to continuously changing perspectives.

{Existing multi-spectral fusion methods \cite{zhang2020multispectral,azarang2019convolutional} attempt to address these issues, but they often rely on rigid calibration or alignment techniques that do not adapt well to the dynamic nature of UAV platforms. Some methods, such as multi-scale context aggregation \cite{yu2015multi} and transformer- or graph-based global reasoning \cite{chen2019graph,luo2022evaluating}, offer partial solutions but fail to handle the extreme viewpoints and rapid changes in UAV trajectories. Moreover, the lack of consistent semantic mapping between RGB and thermal images often results in incomplete or misaligned segmentation outputs, especially for small and complex targets.}

Despite recent advances in UAV-based multi-spectral video segmentation, two key challenges remain insufficiently addressed:
(1) Modal Fusion Dilemma: How to suppress redundant or noisy differences between RGB and thermal modalities, given their inherently distinct characteristics, and effectively extract truly complementary information for robust segmentation;
(2) Temporal Variation Challenge: The high-speed motion of UAV platforms causes rapid changes in imaging parameters such as viewing angle and altitude over time, leading to significant frame-to-frame variations in object appearance within multi-spectral video streams. This is especially problematic for complex structures and small objects. How to effectively model temporal information and extract temporally consistent features of targets remains a crucial issue for achieving accurate segmentation.
These issues highlight the need for more effective solutions to resolve the fusion dilemma and address the dynamic challenges posed by UAV platforms.

The core challenges in UAV-based multi-spectral video segmentation lie in two aspects: (1) effectively extracting complementary information from heterogeneous RGB and thermal modalities while suppressing modality-specific noise, and (2) maintaining semantic consistency across frames under rapid UAV motion and viewpoint shifts.
To address the limitations, this study proposes MSTF-Net, a multi-stage fusion network composed of two key modules: Modality Spatial Complementary Suppression and Enhancement (MSCSE) and Multi-scale Temporal Cross-modality Semantic Consistency (MTCSC).

In the MSCSE module, a cross-modal instance queries is firstly constructed to unify semantic representations from RGB and thermal features. Then, a residual-guided suppression mechanism is introduced to detect and attenuate modality discrepancy noise, which means features present in only one modality without semantic support in the other. By computing query residuals and verifying them with cross-modal similarity, MSCSE selectively suppresses misleading signals while preserving semantically complementary cues. Additionally, a multi-modal spatial consistency constraint ensures that the spatial attention regions of each query are semantically aligned across modalities, enhancing informative feature retention.

In the MTCSC module, this study focuses on temporal variation caused by UAV motion. A granularity-aware temporal fusion strategy dynamically adjusts the temporal receptive field for each frame: nearby frames are processed with fine-grained semantics to retain spatial precision, while distant frames are encoded with coarse granularity to capture global trends and reduce noise. This design ensures robust temporal consistency and semantic continuity across frames with large appearance variations.
Together, these two modules form a unified framework that mitigates cross-modal fusion noise, enhances temporal stability, and delivers semantically consistent segmentation even under low-visibility conditions and dynamic UAV trajectories.
To summarize, the key contributions of this study are four-fold:

$\bullet$ This study proposes MSTF-Net, a multi-stage framework for UAV-based multi-spectral video segmentation that jointly addresses cross-modal discrepancy and temporal semantic inconsistency, significantly improving segmentation accuracy and robustness under dynamic UAV scenarios.

$\bullet$ A MSCSE module is designed which constructs unified instance queries, filters out modality-specific noise via residual modeling, and preserves semantically complementary information with cross-modal consistency constraints.

$\bullet$ A MTCSC module is introduced to adaptively adjust the temporal receptive field according to frame distance, enabling coarse-to-fine temporal fusion that mitigates spatial misalignment and appearance variation.

$\bullet$ Extensive experiments on public UAV-based RGB-T video segmentation benchmarks demonstrate that MSTF-Net achieves state-of-the-art performance, particularly in challenging conditions such as small objects, motion blur, and modality degradation, showcasing the effectiveness of our fusion strategy.

The remainder of this paper is organized as follows. Section 2 reviews related work. Section 3 describes the details of the proposed network. Section 4 presents the analysis 
and experimental results. Section 5 describes the conclusions and future work.

\section{Related Work}

Semantic segmentation has been extensively explored in computer vision, evolving from early RGB image segmentation to more advanced settings that incorporate multi-spectral inputs and UAV platforms. Existing research can be broadly grouped into two lines: (i) segmentation across diverse modalities, including RGB, thermal, and their multi-spectral fusion, with increasing attention to temporal modeling in videos; and (ii) UAV-based semantic segmentation, which leverages the flexibility of aerial imagery but faces unique challenges such as viewpoint variation, motion-induced distortion, and cross-modal alignment. Together, these strands of work provide the foundation for UAV-based multi-spectral video semantic segmentation, yet key issues, including robust modality fusion, temporal consistency, and resilience under dynamic UAV motion---remain underexplored.

\subsection{Semantic Segmentation in Diverse Modalities}
Semantic segmentation has been extensively studied across various data modalities, each presenting unique challenges and advantages \cite{thisanke2023semantic,yuan2024survey}.
Among them, RGB semantic segmentation (RSS) is the most mature and well-explored 
\cite{zhou2024cross,zhou2024boundary,niu2025exploring,li2025efficient,yan2025sgtc}.
To address the limitations of RGB under challenging illumination conditions, multi-spectral semantic segmentation (MSS) has emerged, aiming to integrate RGB and thermal (infrared) imagery for more robust perception \cite{giakoumoglou2024early,chen2025thermal,li2024smooth,ramos2024multispectral,li2024multispectral}. Beyond standard RGB benchmarks, segmentation and dense-prediction techniques are also central to remote sensing and biomedical imaging, including SAR image segmentation and multi-task recognition-segmentation networks \cite{wang2020deep,wang2021deep,wang2022semi}, as well as fluorescence microscopy restoration and cell-level morphological mapping \cite{li2025volume,guan2025cell}. To strengthen feature representation under challenging conditions, prior studies explore convolutional-transformer hybrids, multi-scale feature attention, and hierarchically designed classifiers \cite{wang2022global,wang2023sar,wang2022recognition,shang2023hdss}, together with robustness-oriented learning for limited, degraded, or open-set data via meta-learning, causal intervention, feature refinement, and generative augmentation \cite{wang2023entropy,wang2025limited,wang2024unveiling,wang2023crucial,wang2023sar1,wang2022sar}. Semantic-level and relational reasoning, such as scene-graph generation, unbiased predicate representation, and cross-modal representation transfer, further enriches structured scene understanding \cite{lili2024panoptic,lili2024domain,li2023biased,zhang2023vpgtrans}, while out-of-distribution detection extends reliability to multimodal and video settings \cite{lili2025dpu,lili2025secure}.
RGB video semantic segmentation (RVSS), on the other hand, has leveraged the temporal continuity in RGB videos to improve per-frame segmentation stability and semantic coherence.  Domain adaptation \cite{mai2024pay}, temporal modeling \cite{grammatikopoulou2024spatio,yao2025event}, and cross-frame attention \cite{guo2024vanishing} have been widely adopted to align and integrate temporal features.  These models achieve promising results in dynamic scenes, especially under consistent lighting.

Bringing together both spectral and temporal cues, multi-spectral video semantic segmentation (MVSS) is a nascent yet promising research direction \cite{ji2024unleashing}.  By jointly exploiting complementary modalities (e.g., RGB and thermal) over time, MVSS has the potential to overcome the limitations of single-modality segmentation, especially in low-light or occluded scenarios.  The MVSeg dataset \cite{ji2023multispectral} represents an early benchmark in this space, introducing a baseline model for fusing RGB-thermal video streams.

In summary, RGB video segmentation is mature but vulnerable to lighting changes. By contrast, multi-spectral temporal methods offer greater robustness yet still demand improved RGB--thermal fusion, attention mechanisms, and spatiotemporal learning for real-world deployment.

\subsection{UAV-based Semantic Segmentation}

Early works in UAV-RGB semantic segmentation primarily adapted ground-level segmentation architectures to aerial images \cite{cheng2024methods}. Researchers have explored multi-scale fusion \cite{xiang2023ctfusenet}, multi-perspective transformer \cite{ji2024pptformer}, and lightweight techniques \cite{behera2024lightweight} to handle the scale variation and viewpoint distortions inherent in UAV imagery.
With the growing demand for reliable UAV perception under challenging conditions, recent works have begun to incorporate multi-spectral inputs, especially thermal imagery, to complement RGB sensing \cite{sun2020fuseseg}. Multi-spectral UAV datasets such as CART \cite{lee2024caltech} and UAVM \cite{li2024multispectral} provide RGB-thermal image pairs with pixel-level annotations. 

For UAV-captured videos, some methods have proposed spatiotemporal models for rice weeds identification \cite{lan2021real}, flood disaster \cite{inthizami2022flood}, or traffic analysis \cite{bisio2022traffic}, but these efforts have largely focused on single-modality (RGB) videos. Recurrent and attention-based temporal models, such as multi-view CNN-LSTM networks, have also proven effective for modeling sequential and multi-view observations \cite{wang2021multiview,wang2020multi}. In contrast, UAV-based multi-spectral video semantic segmentation (MVSS) remains largely unexplored. The only known benchmark effort, MVSeg \cite{ji2023multispectral}, introduces a synthetic dataset and a simple RGB-thermal fusion baseline, which does not explicitly address key challenges such as temporal consistency, cross-modal alignment, or scene-adaptive fusion.

In summary, although UAV semantic segmentation for RGB and multi-spectral images is well developed, multi-spectral video segmentation remains in its infancy. It still struggles with motion induced, cross-modal misalignment, heterogeneous spatiotemporal fusion and real-time UAV perception in challenging conditions.

\section{Proposed Method}

To tackle the challenges of modality discrepancy and temporal inconsistency in UAV-based multi-spectral video segmentation, we propose MSTF-Net, a novel framework that integrates both cross-modal semantic refinement and temporal consistency modeling. 

Specifically, we introduce two core modules, (1) Modality Spatial Complementary Suppression and Enhancement (MSCSE), which constructs unified instance queries from RGB and thermal modalities, and applies residual-guided suppression to filter modality-specific noise while preserving complementary semantic cues;
(2) Multi-scale Temporal Cross-Modality Semantic Consistency (MTCSC), which aggregates long-range temporal context using adaptive perceptual granularity, enabling robust feature alignment under dynamic UAV conditions.

The combination of MSCSE and MTCSC allows MSTF-Net to generate temporally stable, semantically accurate predictions even in low-visibility or fast-changing scenes. The detailed architecture and processing pipeline are described in the following sections.

\subsection{Proposed Framework}
As discussed above, UAV-based multi-spectral video segmentation suffers from two core challenges: (1) {Modal Fusion Dilemma}, where the RGB and thermal modalities exhibit heterogeneous feature representations and modality-specific noise, making it difficult to extract truly complementary information; and (2) {Temporal Variation Challenge}, where fast-changing viewpoints and altitudes in UAV platforms lead to significant inconsistencies in spatial semantics across frames, especially for small and complex objects.

\begin{figure}[htbp]
  \centering
  \includegraphics[width=0.92\linewidth]{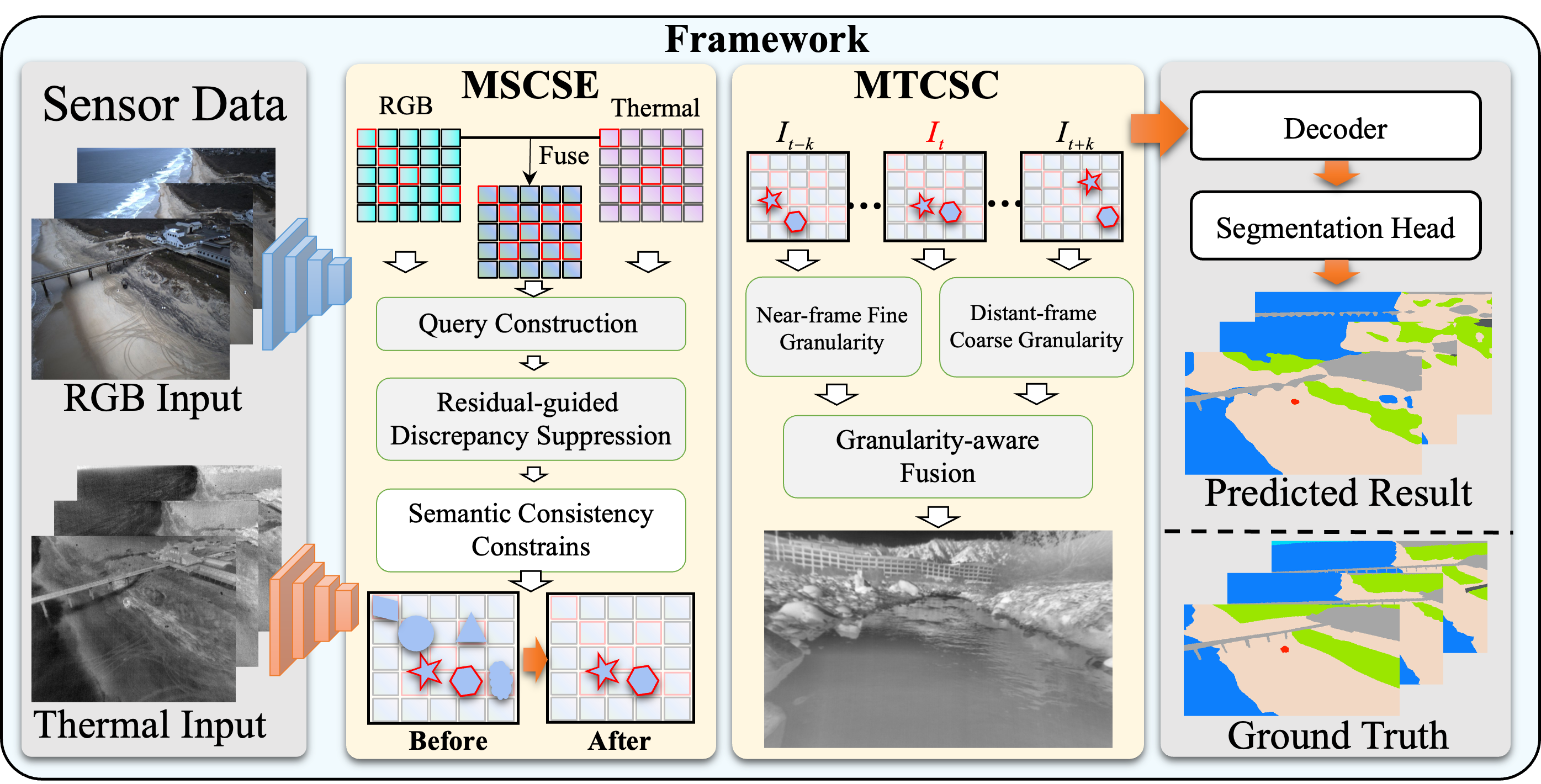}
  \caption{Overview of the proposed MSTF-Net framework for UAV-based multi-spectral video segmentation. The network integrates two key modules: (1) Modality Spatial Complementary Suppression and Enhancement (MSCSE), which constructs unified cross-modal instance queries, suppresses modality-specific discrepancy noise via residual-guided suppression, and enforces spatial consistency across modalities; and (2) Multi-scale Temporal Cross-modality Semantic Consistency (MTCSC), which applies a granularity-aware temporal fusion strategy to adaptively balance fine-grained precision and coarse-grained robustness across varying frame distances. Together, these modules mitigate fusion noise, enhance temporal stability, and ensure semantically consistent segmentation under dynamic UAV motion and challenging visibility conditions.}
  \label{fig:method}
\end{figure}

To address these issues, we propose MSTF-Net which integrates modality spatial complementary enhancement with multi-scale temporal consistency modeling. MSTF-Net consists of two key modules: MSCSE and MTCSC. The overall architecture is illustrated in Fig.~\ref{fig:method}. The pipeline is described as follows.

Given a video clip with $l$ preceding frames and a target frame, we extract per-frame RGB and thermal features using two backbone networks. These features are fused into a shared representation $F_t$ at each time $t$, resulting in fused multi-spectral features $\{F_{t-c}, \dots, F_{t-c_l}, F_t\} \in \mathbb{R}^{h \times w \times c}$. These serve as the input to the MSTF-Net pipeline.

To suppress modality discrepancy noise and preserve semantically aligned, complementary information, we first process the fused features of each frame. At this stage, we perform cross-modal instance query construction using a bi-directional attention mechanism.
To filter modality-specific noise, we compute the residual between modality-specific queries and the fused query, estimate the semantic uncertainty using cross-similarity, and apply confidence-weighted suppression.
The refined queries $\hat{Q}^{rgb}, \hat{Q}^{thermal}$ are then fused into a denoised final query $\hat{Q}^{fusion}$ via uncertainty-aware weighted combination.

After enhancing each frame's fused features, we employ the MTCSC module to model long-range temporal dependencies with adaptive granularity. Specifically, we divide the current target frame $F_t$ into fine-grained windows and reference frames $\{F_{t-c}, \dots, F_{t-c_l}\}$ into coarser windows depending on their distance from $t$.
We then extract context tokens $c_{i,j}$ from each reference frame $F'_j$ for every window $F'_t[i]$ in the target frame. These are concatenated across time.
This fusion enables each target query to leverage both temporally local precision and temporally global consistency, forming a coherent and stable semantic representation.

Overall, the proposed {MSTF-Net} addresses the two key challenges in UAV-based multi-spectral video segmentation. The {MSCSE} module suppresses modality discrepancy noise and enhances complementary semantics, while the {MTCSC} module models multi-scale temporal context to mitigate appearance inconsistencies led by fast-moving UAV. Together, they enable robust and accurate segmentation under dynamic and low-visibility UAV conditions.

\subsection{Modality Spatial Complementary Suppression and Enhancement}

Multi-spectral inputs such as RGB and thermal data provide complementary visual cues, yet they also suffer from modality-specific noise due to inherent sensing differences. Information that exists in one modality but lacks support in the other often leads to semantic misalignment or hallucinated regions in fused features, commonly referred to as {modality discrepancy noise}. To address this, we propose the MSCSE module, which aims to extract semantically aligned, complementary information across modalities while suppressing unsupported noisy signals. MSCSE consists of three stages: instance query construction, semantic consistency enhancement, and residual-guided discrepancy suppression.

The MSCSE module consists of two steps and two constrains, the two steps are 
1) Cross-modal Instance Query Construction for unified representation, 2) Residual-guided Discrepancy Suppression to suppress modality discrepancy noise.
The details are described as follows.

Stage 1: Cross-modal Instance Query Construction.
Given visual features from RGB and thermal branches, denoted as $F_{rgb}, F_{thermal} \in \mathbb{R}^{h \times w \times c}$, we first project them into a shared feature space:
\begin{equation}
F'_{rgb} = \phi_{rgb}(F_{rgb}), \quad F'_{thermal} = \phi_{thermal}(F_{thermal}),
\end{equation}
We then obtain a fused feature representation and generate instance queries via bi-directional cross-attention and decoding:
\begin{align}
F_{fused} &= \text{Fusion}(F'_{rgb}, F'_{thermal}),\\
Q_{rgb} &= \text{Decoder}(\text{BiAtt}(F'_{rgb}, F'_{thermal})), \\
Q_{thermal} &= \text{Decoder}(\text{BiAtt}(F'_{thermal}, F'_{rgb})), \\
Q_{fused} &= \text{Decoder}(F_{fused}).
\end{align}
where, $Q_{rgb}, Q_{thermal}, Q_{fused} \in \mathbb{R}^{N \times C}$ represent instance-level queries carrying modality-specific and fused semantic cues.

Stage 2: Residual-guided Discrepancy Suppression.
we propose a residual-guided suppression mechanism to explicitly compress modality discrepancy noise. This type of noise refers to information that appears in one modality but lacks semantic support in the other, often manifesting as blurred boundaries, spatial misalignments, or false activations such as thermal hotspots. These signals typically lack cross-modal consistency and are therefore unlikely to contribute meaningful complementary information.
To suppress modality discrepancy noise, we first compute the residual representation between each modality-specific query and the fused query as::
\begin{equation}
\begin{aligned}
\mathbf{R}^{rgb} &= \mathbf{Q}^{rgb} - \mathbf{Q}^{fusion}, \\
\mathbf{R}^{thermal} &= \mathbf{Q}^{thermal} - \mathbf{Q}^{fusion}.
\end{aligned}
\end{equation}
These residuals encode modality-specific components that may include inconsistent or noisy features not shared across modalities. To determine whether the residual contains semantically meaningful information, we compute its similarity with the query features from the other modality:

\begin{equation}
\begin{aligned}
s^{rgb \rightarrow thermal} &= \text{Sim}\left( \mathbf{R}^{rgb}, \mathbf{Q}^{thermal} \right), \\
s^{thermal \rightarrow rgb} &= \text{Sim}\left( \mathbf{R}^{thermal}, \mathbf{Q}^{rgb} \right),
\end{aligned}
\end{equation}
where $\text{Sim}(\cdot)$ denotes a similarity function, we use cosine similarity here. Lower similarity indicates that the residual likely lacks semantic support in the other modality and is thus more likely to be noise.
Thus, We convert the similarity scores into uncertainty weights that guide the degree of suppression via:

\begin{equation}
\begin{aligned}
\beta^{rgb} &= 1 - \sigma\left( s^{rgb \rightarrow thermal} \right), \\
\beta^{thermal} &= 1 - \sigma\left( s^{thermal \rightarrow rgb} \right),
\end{aligned}
\end{equation}
where $\sigma(\cdot)$ is the sigmoid function. The final query representations after residual suppression are given by:

\begin{equation}
\begin{aligned}
\hat{\mathbf{Q}}^{rgb} &= \mathbf{Q}^{rgb} - \beta^{rgb} \cdot \mathbf{R}^{rgb}, \\
\hat{\mathbf{Q}}^{thermal} &= \mathbf{Q}^{thermal} - \beta^{thermal} \cdot \mathbf{R}^{thermal}.
\end{aligned}
\end{equation}
This uncertainty-aware gating mechanism enables soft suppression of modality-specific noise while preserving semantically supported residuals.
After noise suppression, we obtain the refined modality-specific queries $\hat{\mathbf{Q}}^{rgb}$ and $\hat{\mathbf{Q}}^{thermal}$. To obtain the final fused query $\mathbf{Q}^{fusion}$, we perform uncertainty-aware weighted fusion as:
\begin{equation}
\hat{\mathbf{Q}^{fusion}} = \alpha^{rgb} \cdot \hat{\mathbf{Q}}^{rgb} + \alpha^{thermal} \cdot \hat{\mathbf{Q}}^{thermal},
\end{equation}
where the weights $\alpha^{rgb}$ and $\alpha^{thermal}$ are normalized from the confidence scores derived from the suppression step, $\alpha^{rgb} = \frac{1 - \beta^{rgb}}{(1 - \beta^{rgb}) + (1 - \beta^{thermal})}$, $\alpha^{thermal} = \frac{1 - \beta^{thermal}}{(1 - \beta^{rgb}) + (1 - \beta^{thermal})}$.
As a result, the final queries are purer, more discriminative, and more reliable for downstream tasks.

Constrains for Semantic Consistency and Informative Representation. 
To encourage consistent spatial semantics across modalities, we enforce a {multi-modal spatial consistency constraint} by aligning the attention activation maps $A^i \in \mathbb{R}^{H \times W}$ of each query token $q^i$:
\begin{align}
\mathcal{L}_{align} = \sum_{i=1}^N \| A^i_{rgb} - A^i_{thermal} \|_2^2.
\end{align}
This multi-modal spatial consistency constraint, which enforces that the spatial activation regions of corresponding query vectors should be consistent across different modalities, while the activation regions of different query vectors should remain distinct both within and across modalities. This constraint encourages the query vectors to preserve meaningful target information by leveraging cross-modal spatial consistency.


In summary, MSCSE constructs unified cross-modal queries, preserves modality-aware semantics, and explicitly suppresses modality discrepancy noise through residual modeling and mutual verification, enabling semantically coherent and noise-resilient multi-modal representations.

\subsection{Multi-scale Temporal Cross-Modality Semantic Consistency}

In multi-spectral video segmentation, fusing temporal information from past frames is crucial for capturing dynamic object semantics and improving robustness under noisy or ambiguous conditions. However, features from distant frames often exhibit spatial misalignment and weaker relevance to the current frame. To address this, we propose a MTCSC module that aggregates contextual information from multi-temporal fused features with varying perceptual granularity: frames closer to the current frame are processed with finer granularity and local attention, while distant frames adopt coarser granularity and broader receptive fields to reduce noise and highlight stable semantic patterns.

The details of the MTCSC module is as follows.
Let $t$ denote the current frame, and $\{t-c, \dots, t-c_l\}$ be its preceding $l$ reference frames. We denote the fused features from RGB and thermal inputs at each frame as $\{F_{t-c}, \dots, F_{t-c_l}, F_t\} \in \mathbb{R}^{h \times w \times c}$ for simplicity.

We define a temporal index set $U = \{t-c, \dots, t-c_l, t\}$ and process each frame's fused features $F_j$ with window-based partitioning. Specifically, $F_t$ is divided into fine-grained windows of size $s \times s$, and distant frame features are partitioned using coarser window sizes depending on their temporal distance to $t$. This results in a set of token windows for each $F_j$, denoted as $F'_j$:
\begin{align}
F_j \rightarrow F'_j \in \mathbb{R}^{\frac{h}{s_j} \times \frac{w}{s_j} \times s_j^2 \times c},
\end{align}
where $s_j$ is the window size, increasing with frame distance from $t$.

For each token window $F'_t[i]$ from the target frame, we extract contextual tokens from $F'_j$ of each reference frame $j \in U$ by retrieving $r_j \times r_j$ neighboring tokens (where $r_j$ is a receptive field size increasing with temporal distance). These context tokens are then projected by $ F'_j \rightarrow E_j \in \mathbb{R}^{\frac{h}{s_j} \times \frac{w}{s_j} \times c}$.
$c_{i,j}$ denote the context tokens retrieved from $E_j$ for window $i$ in $F_t$. The final context representation for window $i$ is:
\begin{align}
c_i = \text{Concat}(\{c_{i,j} \mid j \in U\}).
\end{align}

The concatenated context $c_i$ is fused with the corresponding query in the target frame to refine temporal consistency:
\begin{align}
\tilde{q}_i = \text{Fusion}(q_i, c_i),
\end{align}
where $q_i$ is the query token from $F_t$. This fusion allows both fine-grained and coarse-grained context from surrounding frames to modulate and enhance the semantic representation of the current frame.

In summary, MTCSC constructs a hierarchical temporal fusion mechanism where fused features from nearby frames provide detailed local support, while distant frames contribute coarse but globally consistent semantics. This multi-scale temporal modeling enables better alignment across time and improves both static and motional context consistency in multi-spectral segmentation.

\section{Experiments}

\begin{figure}[htbp]
  \centering
  \includegraphics[width=0.48\linewidth]{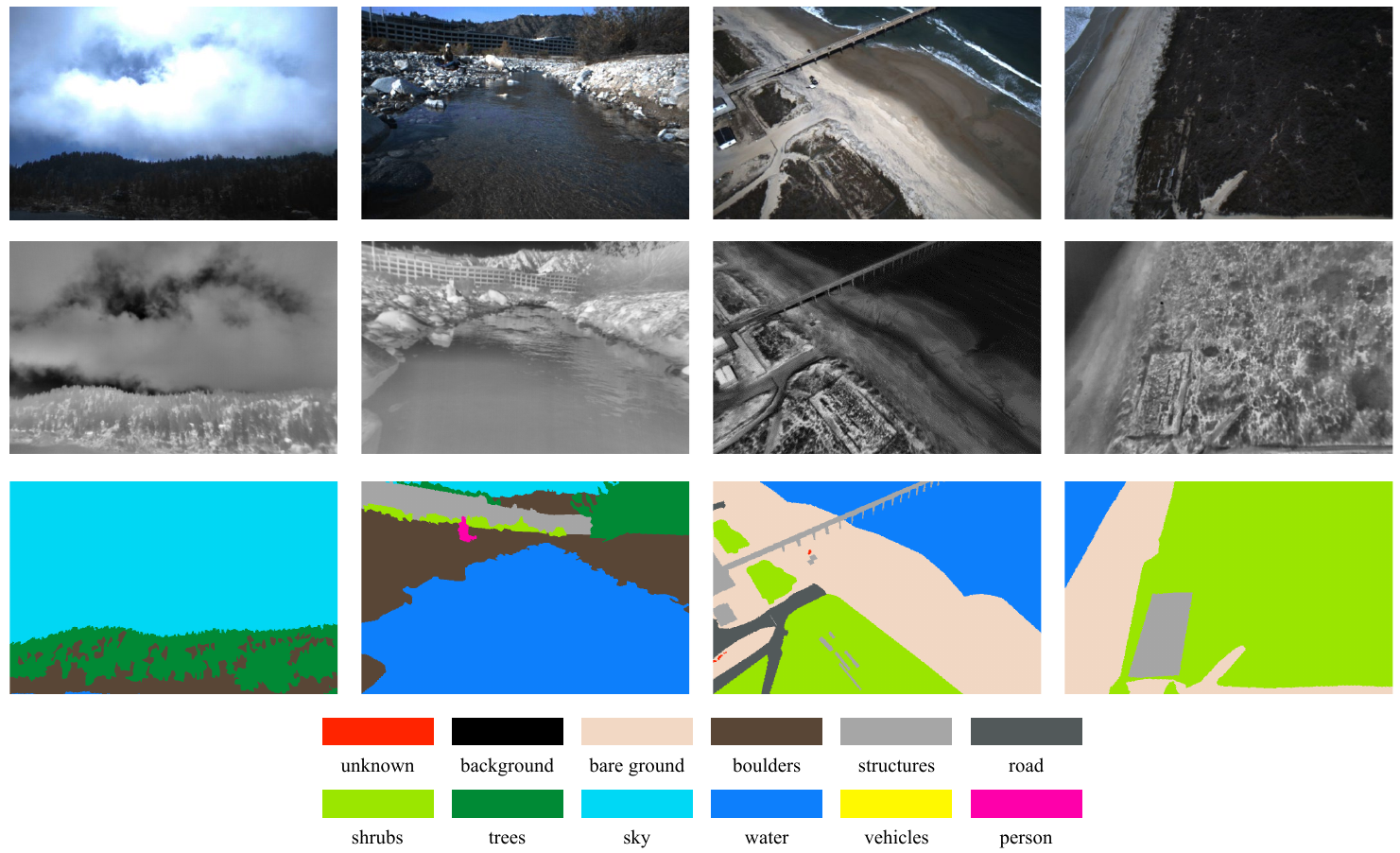}
  \caption{Overview of the Caltech Aerial RGB--Thermal (CART) dataset, covering 37 UAV trajectories over diverse terrains with paired RGB and 16-bit thermal images. A subset of 4,195 frames is annotated into classes such as water, sky, structures, vehicles, and persons, with multiple splits (general, temporal, terrain, region) for evaluating model generalization under varying conditions.}
  \label{fig:dataset_cart}
\end{figure}

\begin{figure}[htbp]
  \centering
  \includegraphics[width=0.48\linewidth]{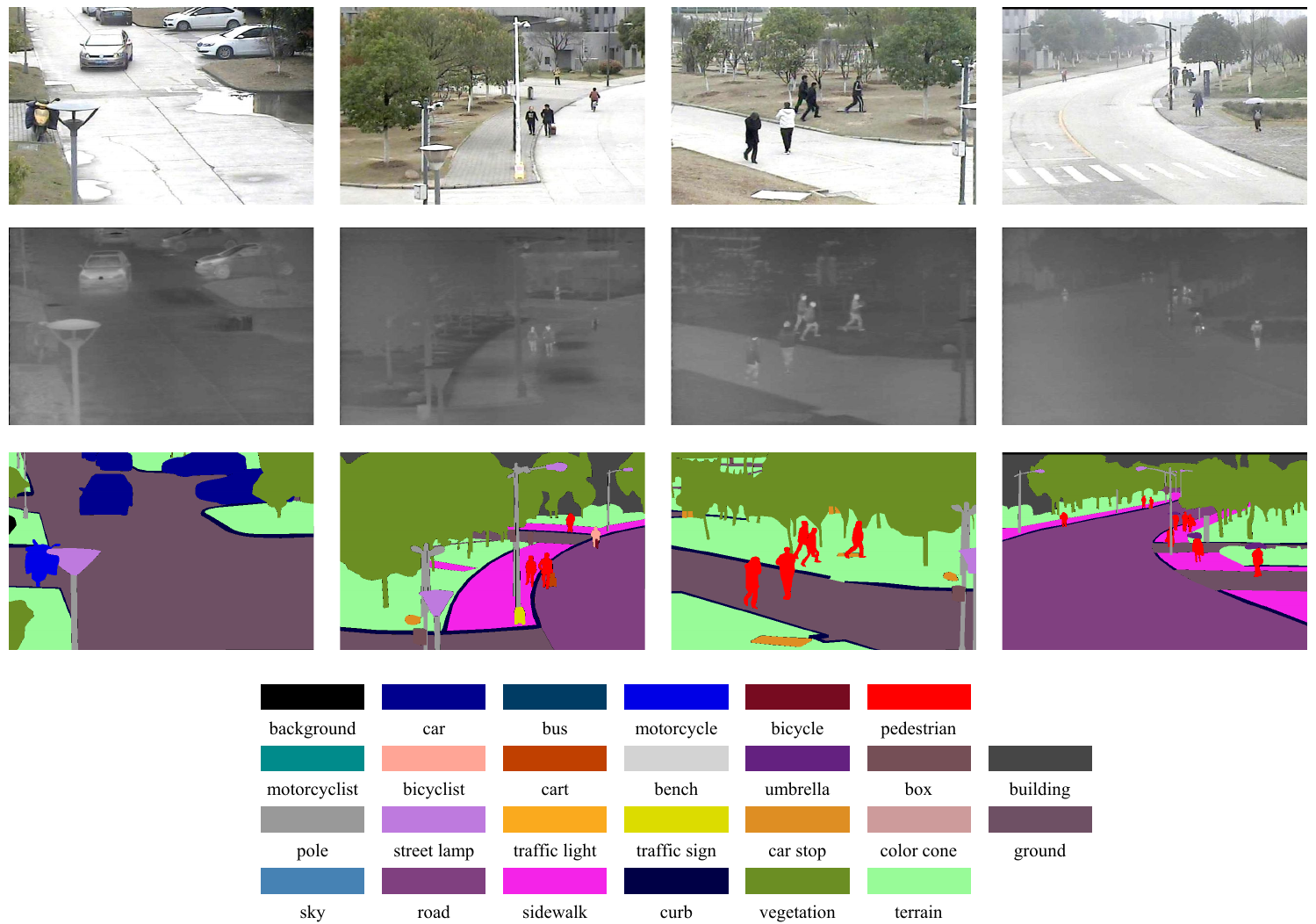}
  \caption{Overview of the MVSeg dataset for multi-spectral video semantic segmentation. It consists of 738 calibrated RGB--thermal video clips, with 3,545 frames densely annotated into 26 semantic categories. Captured under diverse conditions (daytime, nighttime, rain, snow, and low-light), each clip lasts ~5 seconds at 15 fps. The dataset achieves a 98.96\% dense labeling rate.}
  \label{fig:dataset_MVSeg}
\end{figure}

This section presents a comprehensive evaluation of the proposed MSTF-Net. We begin by introducing the datasets and implementation details, including CART for aerial RGB--thermal imagery and MVSeg for multi-spectral video segmentation (Figs.~\ref{fig:dataset_cart} and \ref{fig:dataset_MVSeg}). Next, we conduct quantitative comparisons with a wide range of state-of-the-art methods to assess accuracy, efficiency, and robustness across diverse conditions. We then provide qualitative analyses to illustrate the effectiveness of our approach in challenging UAV scenarios such as small objects, motion blur, and adverse illumination. Finally, we perform detailed ablation studies to investigate the contribution of each component, including fusion strategy, MSCSE, and MTCSC---across different backbones, offering insights into their individual and combined impact.

\subsection{Datasets and Implementation Details}
The datasets used in the experiments include Caltech Aerial RGB-Thermal (CART) dataset \cite{lee2024caltech} and MVSeg dataset \cite{ji2023multispectral}.

The Caltech Aerial RGB-Thermal (CART) dataset contains $37$ synchronized trajectories across lakes, shrubs, coasts, deserts, and mountains, providing paired RGB and 16-bit thermal images. A subset of $4195$ paired RGB and thermal images are semantically annotated into classes such as water, sky, structures, vehicles, and persons. The dataset is partitioned into general (75/12.5/12.5 train/val/test), temporal (twilight/day/night), terrain-based, and region-based splits for assessing the performance and generalization of multi-spectral models across various illumination conditions.


The benchmark uses MVSeg dataset, which is a comprehensive dataset designed for multi-spectral video semantic segmentation. MVSeg comprises 738 calibrated RGB--thermal video clips, of which 3545 frames have been densely annotated with pixel-level semantic labels spanning 26 object and scene categories (e.g., vehicles, people, buildings, vegetation). Captured under diverse and challenging conditions, including daytime, nighttime, rain, snow, and other low-light scenarios, each clip runs for approximately five seconds at 15 fps. Rigorous annotation protocols ensure a dense pixel-wise labeling rate of 98.96\%, far surpassing previous multi-spectral benchmarks. The dataset is split into 452 training, 84 validation, and 202 test videos, with the test set further subdivided into daytime and nighttime subsets to facilitate detailed robustness evaluation.

Model training was conducted on both CART dataset and MVSeg dataset, For the CART dataset, the images are uniformly resized to 300$\times$480, and for the MVSeg dataset, the images are resize to 320$\times$480, then performed random horizontal flipping and cropping to avoid the over-fitting. Training procedures employed the Adam optimizer with an initial learning rate of 2e-4, weight decay of 0.1 each 100 iterations, and the Adam optimizer variant for robust gradient updates. The proposed method and comparison models were all trained and evaluated using 2 NVIDIA A100 GPUs.



\subsection{Quantitative Comparison with the State-of-the-Arts}

\begin{table}[ht]
  \centering
  \setlength{\tabcolsep}{8pt}
  \renewcommand{\arraystretch}{1.2}

  \caption{Quantitative evaluation on the test set of the CART dataset}
  \label{tab:metrics1}
  \begin{adjustbox}{max width=\textwidth}
    \begin{tabular}{l l c c c c c c c}
      \toprule\toprule
      Methods                    & Backbone   & Resolution  & Mem (G) & Time (ms)
        & Dice   & FwIoU  & mIoU (\%) & Acc \\
      \midrule
      CCNet \cite{huang2019ccnet}
        & ResNet-50       & $300\times480$& 4.7  & 9.7 
        & 60.98  & 71.46  & 43.86    & 60.22  \\
      OCRNet \cite{yuan2020object}
        & ResNet-50       & $300\times480$& 4.6  & 9.3 
        & 61.12  & 72.25  & 43.98    & 59.69  \\
      STM \cite{oh2019video}
        & ResNet-50       & $300\times480$& 20.6 & 10.6
        & 61.33  & 70.12  & 44.23    & 58.83  \\
      LMANet \cite{paul2021local}
        & ResNet-50       & $300\times480$& 9.9  & 9.2 
        & 61.72  & 71.17  & 44.65    & 59.72  \\
      MFNet \cite{ha2017mfnet}
        & Mini-inception  & $300\times480$& \textbf{2.2}  & \textbf{5.1} 
        & 60.86  & 69.67  & 43.72    & 58.91  \\
      RTFNet \cite{sun2019rtfnet}
        & ResNet-152      & $300\times480$& 8.1  & 44.8
        & 61.86  & 70.65  & 44.78    & 59.60  \\
      EGFNet \cite{zhou2022edge}
        & ResNet-152      & $300\times480$& 6.0  & 71.4
        & 62.98  & 72.14  & 45.97    & 62.67  \\
      \midrule
      FCN \cite{long2015fully}
        & ResNet-50       & $300\times480$& 4.5  & 6.2 
        & 59.84  & 71.94  & 42.68    & 60.25  \\
      MVNet(FCN) \cite{ji2023multispectral}
        & ResNet-50       & $300\times480$& 16.9 & 14.5
        & 63.32  & 73.90  & 46.29    & 62.91  \\
      PSPNet \cite{zhao2017pyramid}
        & ResNet-50       & $300\times480$& 4.7  & 9.3 
        & 60.26  & 71.11  & 43.12    & 59.54  \\
      MVNet(PSPNet) \cite{ji2023multispectral}
        & ResNet-50       & $300\times480$& 10.6 & 20.5
        & 63.87  & 73.28  & 46.93    & 61.78  \\
      DeepLabv3+ \cite{chen2018encoder}
        & ResNet-50       & $300\times480$& 4.4 & 7.8 
        & 60.37  & 70.55  & 43.23    & 58.33  \\
      MVNet(DeepLabv3+) \cite{ji2023multispectral}
        & ResNet-50       & $300\times480$& 17.7 & 17.5
        & 63.98  & 73.72  & 47.03    & 62.13  \\
      Ours (MSTF-Net)
        & ResNet-50       & $300\times480$& 18.2 & 18.7
        & \textbf{68.24}  & \textbf{78.75}  & \textbf{51.80}    & \textbf{66.15}  \\
      \bottomrule\bottomrule
    \end{tabular}
  \end{adjustbox}
\end{table}

\begin{table}[ht]
  \centering
  \normalsize
  \setlength{\tabcolsep}{10pt}
  \renewcommand{\arraystretch}{1.5}

  \caption{The quantitative results on each class of the CART benchmark test set (Part I)}
  \label{tab:CART-per-class-miou-part1}

  \begin{adjustbox}{max width=\textwidth}
    \begin{tabular}{l *{7}{c}}
      \toprule\toprule
      & CCNet\cite{huang2019ccnet}
      & OCRNet\cite{yuan2020object}
      & STM\cite{oh2019video}
      & LMANet\cite{paul2021local}
      & MFNet \cite{ha2017mfnet}
      & RTFNe \cite{sun2019rtfnet}
      & EGFNet \cite{zhou2022edge} \\
      \midrule
      Bare ground & 42.83 & 42.52 & 43.18 & 43.12 & 42.70 & 43.30 & \textbf{44.98} \\
      Boulders    & 86.74 & 86.46 & 87.29 & \textbf{89.35} & 86.52 & 87.32 & 89.10 \\
      Structures  & 39.03 & 38.77 & 39.48 & 39.21 & 38.80 & 39.34 & \textbf{40.79} \\
      Road        & 42.63 & 43.08 & 42.86 & 43.62 & 42.27 & 43.80 & \textbf{44.60} \\
      Shrubs      & 35.24 & 34.91 & 35.53 & 35.29 & 34.98 & 35.28 & \textbf{36.84} \\
      Trees       & 38.68 & 39.20 & 38.91 & 39.84 & 38.59 & 39.79 & \textbf{40.65} \\
      Sky         & 81.23 & 79.82 & 80.59 & 80.47 & 79.61 & 81.12 & \textbf{81.33} \\
      Water       & 80.12 & 79.85 & 80.60 & 80.12 & 79.68 & \textbf{81.31} & 81.22 \\
      Vehicles    & 35.02 & 34.89 & 35.47 & 35.19 & 35.02 & 35.35 & \textbf{37.02} \\
      Person      & 23.20 & 23.56 & 23.16 & 23.85 & 22.85 & 24.06 & \textbf{24.22} \\
      \midrule
      Mean IoU (\%) & 43.86 & 43.98 & 44.23 & 44.65 & 43.72 & 44.78 & \textbf{45.97} \\
      \bottomrule\bottomrule
    \end{tabular}
  \end{adjustbox}
\end{table}

\begin{table}[ht]
  \centering
  \normalsize
  \setlength{\tabcolsep}{10pt}
  \renewcommand{\arraystretch}{1.5}

  \caption{The quantitative results on each class of the CART benchmark test set (Part II)}
  \label{tab:CART-per-class-miou-part2}

  \begin{adjustbox}{max width=\textwidth}
    \begin{tabular}{l *{7}{c}}
      \toprule\toprule
      & FCN\cite{long2015fully}
      & MVNet (FCN)
      & PSPNet \cite{zhao2017pyramid}
      & MVNet (PSPNet)
      & DeepLabv3+\cite{chen2018encoder}
      & MVNet (DeepLabv3+)
      & MSTF-Net \\
      \midrule
      Bare ground & 41.40 & 45.22 & 42.18 & 45.45 & 41.90 & 45.53 & \textbf{50.60} \\
      Boulders    & 85.13 & 89.41 & 85.64 & 90.16 & 85.69 & \textbf{91.56} & 90.91 \\
      Structures  & 37.59 & 41.26 & 37.93 & 41.72 & 38.12 & 41.88 & \textbf{45.52} \\
      Road        & 41.66 & 44.82 & 42.16 & 45.38 & 41.83 & 46.13 & \textbf{50.15} \\
      Shrubs      & 33.86 & 37.06 & 34.12 & 37.34 & 34.25 & 37.80 & \textbf{40.86} \\
      Trees       & 38.01 & 40.83 & 38.45 & 41.35 & 37.98 & 41.77 & \textbf{45.89} \\
      Sky         & 79.90 & 80.76 & 79.54 & 81.47 & 79.70 & 80.30 & \textbf{81.88} \\
      Water       & 79.94 & 80.77 & 79.43 & \textbf{81.55} & 79.70 & 80.88 & 81.46 \\
      Vehicles    & 33.84 & 37.15 & 34.38 & 37.12 & 34.09 & 37.68 & \textbf{41.14} \\
      Person      & 22.89 & 24.38 & 23.20 & 24.77 & 22.66 & 25.15 & \textbf{27.41} \\
      \midrule
      Mean IoU (\%) & 42.68 & 46.29 & 43.12 & 46.93 & 43.23 & 47.03 & \textbf{51.80} \\
      \bottomrule\bottomrule
    \end{tabular}
  \end{adjustbox}
\end{table}

\textbf{Results on CART Dataset.}
Table~\ref{tab:metrics1} reports the quantitative comparison results on the CART test set. Our method outperforms all state-of-the-art baselines across all three metrics. Specifically, it achieves a Dice score of 68.24, a FwIoU of 78.75, and a mIoU of 51.80\%, surpassing the previous best (MVNet based on DeepLabv3+) by +4.26\% in Dice and +4.1\% in mIoU. Compared to single-modality baselines such as DeepLabv3+ and PSPNet, our model significantly boosts semantic segmentation performance, demonstrating the benefit of enhanced RGB-T fusion and attentive decoding. Notably, despite slightly increased memory consumption (18.2G), our method maintains a reasonable inference speed (18.7 ms), confirming its efficiency-accuracy trade-off for real-world deployment. Furthermore, per-class IoU results in Tables~\ref{tab:CART-per-class-miou-part1} and~\ref{tab:CART-per-class-miou-part2} show that MSTF-Net achieves the highest IoU on all 10 semantic categories. The most pronounced gains occur on Bare ground (+5.07\% over MVNet(DeepLabV3+)), Road (+4.02\%), Trees (+4.12\%), and Structures (+3.64\%), leading to a mean IoU improvement of +4.77\%. These consistent per-class enhancements validate the robustness of our multi-scale spatiotemporal feature fusion and cross-modal attention mechanisms across diverse scene elements. Overall, MSTF-Net establishes a new state of the art on the CART benchmark, offering an excellent balance between segmentation accuracy and real-time inference efficiency.

\begin{table}[ht]
  \centering
  \scriptsize
    \normalsize
  \setlength{\tabcolsep}{10pt}
  \renewcommand{\arraystretch}{1.5}
  \caption{More quantitative evaluation on the test set of the MVSeg dataset}
  \label{tab:metrics2}
  \begin{adjustbox}{width=\textwidth}
  \begin{tabular}{l|l|l|c|c|c|c|c|c|c}
    \toprule \toprule
    Methods                 &Model Setup   & Backbone   & Resolution & Param (M) & Mem (G) & Time (ms) & Dice & FwIoU  & mIoU (\%) \\
    \midrule
    CCNet \cite{huang2019ccnet}       & RGB-T & ResNet-50  & $320\times480$ & 52.3 & 4.9 & 10.3 & 68.16 &  72.36  & 51.70 \\
    OCRNet \cite{yuan2020object}     & RGB-T & ResNet-50  & $320\times480$ & 43.6 & 4.8 & 9.8 & 68.74 &  72.19  & 52.38 \\
    STM \cite{oh2019video}& RGB-T & ResNet-50  & $320\times480$ & 44.1 & 21.9 & 11.2 & 68.87 & 70.88  & 52.51 \\
    LMANet \cite{paul2021local}
                               & RGB-T & ResNet-50  & $320\times480$ & 44.1 & 10.4 & 9.7 &69.06 & 73.17  & 52.73 \\
    MFNet \cite{ha2017mfnet}
                               & RGB-T & Mini-inception & $320\times480$ & \textbf{0.71} & \textbf{2.3} & \textbf{5.3} & 68.10 &  72.67  & 51.63 \\
    RTFNet \cite{sun2019rtfnet}
                               & RGB-T & ResNet-152 & $320\times480$ & 254.5 & 8.6 & 47.7 & 69.10 &  73.65  & 52.77 \\
    EGFNet \cite{zhou2022edge}
                               & RGB-T & ResNet-152 & $320\times480$ & 123.2 & 6.3 & 75.8 & 69.66 &  73.02  & 53.44 \\
    \midrule
    FCN \cite{long2015fully}
                              & RGB-T  & ResNet-50 & $320\times480$ &24.1 &4.7 &6.5 & 67.26 &  71.94  & 50.67 \\
    MVNet(FCN) \cite{ji2023multispectral}
                               & RGB-T & ResNet-50 & $320\times480$ & 50.2&17.9 & 15.3& 70.01 &  73.90  & 53.90 \\
    PSPNet \cite{zhao2017pyramid}
                               & RGB-T & ResNet-50 & $320\times480$ & 47.9&5.0 &9.8 & 67.88 &  71.11  & 51.38 \\
    MVNet(PSPNet) \cite{ji2023multispectral}
                               & RGB-T & ResNet-50 & $320\times480$ &98.3& 11.2&21.3 & 70.47 &  73.28  & 54.36 \\

    \multirow{2}{*}{DeepLabv3+ \cite{chen2018encoder}}
                               & RGB &ResNet-50 & $320\times480$  & 19.2& 2.9 & 5.3&66.72 &  73.66  & 50.04 \\
                               & RGB-T &ResNet-50 & $320\times480$ & 41.6 & 4.6 & 8.1 &  68.07  & 74.10&51.59 \\

     \multirow{2}{*}{MVNet (DeepLabv3+) \cite{ji2023multispectral}}
                               & RGB  & ResNet-50 & $320\times480$ &42.3 & 4.7 & 8.7&  67.33  & 73.93&51.21 \\
                               & RGB-T & ResNet-50  & $320\times480$ & 88.4 & 18.8 & 18.4 &  70.64  & 76.67 &54.52 \\
                               \midrule
     \multirow{1}{*}{Ours (MSTF-Net)}
                              
                               & RGB-T & ResNet-50 & $320\times480$ &92.1 &19.5 &20.1 &\textbf{72.14} & \textbf{78.32}  & \textbf{56.42}  \\
    \bottomrule \bottomrule
  \end{tabular}
  \end{adjustbox}
\end{table}

\begin{table}[ht]
  \centering
  \caption{The quantitative results on each class of the MVSeg benchmark test set (Part I)}
  \label{tab:per-class-miou-part1-mvseg}
  \begin{adjustbox}{max width=0.95\textwidth}
    \begin{tabular}{l*{7}{c}}
      \toprule\toprule
      & CCNet\cite{huang2019ccnet}
      & OCRNet\cite{yuan2020object}
      & STM\cite{oh2019video}
      & LMANet\cite{paul2021local}
      & MFNet \cite{ha2017mfnet}
      & RTFNe \cite{sun2019rtfnet}
      & EGFNet \cite{zhou2022edge} \\
\midrule
      Background    & 38.40 & \textbf{40.39} & 39.13 & 36.63 & 36.21 & 34.31 & 34.48 \\
      Car           & 79.75 & \textbf{81.82} & 79.59 & 81.35 & 80.54 & 81.43 & 80.77 \\
      Bus           & 36.82 & 37.53 & 34.59 & 36.48 & 40.03 & 35.66 & \textbf{41.61} \\
      Motorcycle    & 29.78 & 36.02 & 32.94 & \textbf{36.57} & 31.59 & 32.34 & 33.41 \\
      Bicycle       & 60.04 & \textbf{62.60} & 60.17 & 57.67 & 53.89 & 55.12 & 61.41 \\
      Pedestrian    & 55.00 & 55.59 & 54.05 & 58.95 & 58.04 & \textbf{61.76} & 61.75 \\
      Motorcyclist  &  4.10 &  4.43 &  3.33 & 12.23 &  6.22 & \textbf{13.83} & 11.88 \\
      Bicyclist     & 27.21 & 25.48 & 25.52 & \textbf{32.08} & 26.06 & 30.64 & 31.75 \\
      Cart          & 49.22 & \textbf{58.00} & 54.17 & 46.41 & 49.73 & 50.60 & 54.17 \\
      Bench         & \textbf{59.64} & 53.18 & 58.05 & 51.95 & 51.54 & 55.57 & 54.22 \\
      Umbrella      & 40.72 & 42.97 & 45.68 & 34.32 & 36.07 & \textbf{45.95} & 44.75 \\
      Box           & 36.66 & 37.87 & 38.98 & 39.69 & \textbf{40.11} & 37.83 & 38.79 \\
      Pole          & 51.12 & 50.33 & \textbf{54.12} & 49.01 & 46.31 & 44.39 & 45.57 \\
      Street Lamp   & 49.50 & 56.84 & \textbf{61.41} & 55.45 & 53.50 & 58.11 & 50.74 \\
      Traffic Light & 35.68 & 28.54 & \textbf{40.93} & 36.80 & 36.86 & 37.71 & 35.67 \\
      Traffic Sign  & 35.91 & 37.24 & 43.25 & 41.12 & 37.38 & 36.53 & \textbf{44.11} \\
      Car Stop      & 35.63 & \textbf{37.25} & 32.36 & 33.34 & 34.37 & 33.91 & 35.50 \\
      Color Cone    & 22.56 & 20.12 & 22.07 & \textbf{24.52} & 18.19 & 15.84 & 21.02 \\
      Sky           & 84.31 & 84.29 & 88.25 & 87.40 & 90.53 & \textbf{90.88} & 90.27 \\
      Ground        & 86.43 & 88.04 & 78.97 & \textbf{89.16} & 86.12 & 86.75 & 85.79 \\
      Road          & 90.45 & 89.79 & 88.52 & \textbf{90.99} & 90.85 & 90.73 & 90.70 \\
      Sidewalk      & 54.88 & 56.13 & 51.06 & \textbf{58.90} & 57.75 & 57.07 & 58.15 \\
      Curb          & 49.13 & 45.94 & 49.25 & 48.93 & 46.79 & \textbf{49.35} & 47.81 \\
      Terrain       & 77.37 & 76.69 & 75.24 & 76.27 & 77.94 & 78.67 & \textbf{79.06} \\
      Vegetation    & 78.37 & 78.31 & 78.41 & 77.74 & 79.09 & \textbf{80.24} & 79.92 \\
      Building      & 75.62 & 76.54 & 75.16 & \textbf{77.06} & 76.62 & 76.93 & 76.19 \\
      \midrule
      Mean IoU (\%) & 51.70 & 52.38 & 52.51 & 52.73 & 51.63 & 52.77 & \textbf{53.44} \\
      \bottomrule\bottomrule
    \end{tabular}
  \end{adjustbox}
\end{table}

\begin{table}[ht]
  \centering
  \caption{The quantitative results on each class of the MVSeg benchmark test set (Part II)}
  \label{tab:per-class-miou-part2-mvseg}
  \begin{adjustbox}{max width=0.95\textwidth}
    \begin{tabular}{l*{7}{c}}
      \toprule\toprule
      & FCN\cite{long2015fully}
      & MVNet (FCN)
      & PSPNet \cite{zhao2017pyramid}
      & MVNet (PSPNet)
      & DeepLabv3+\cite{chen2018encoder}
      & MVNet (DeepLabv3+)
      & MSTF-Net \\
      \midrule
      Background    & 36.00 & 35.38 & 34.37 & 39.79 & 35.09 & 39.79 & \textbf{40.33} \\
      Car           & 78.64 & 81.39 & 80.74 & 81.45 & 80.66 & 80.98 & \textbf{82.45} \\
      Bus           & 36.42 & 37.40 & \textbf{39.69} & 38.95 & 36.87 & 31.50 & 37.54 \\
      Motorcycle    & 30.92 & 33.78 & 32.66 & 36.42 & \textbf{37.40} & 35.39 & 35.49 \\
      Bicycle       & 52.56 & 60.70 & 54.71 & 59.59 & 55.24 & 65.34 & \textbf{66.60} \\
      Pedestrian    & 52.36 & 61.63 & 57.39 & 59.18 & 54.46 & 58.30 & \textbf{59.19} \\
      Motorcyclist  &  8.80 & 10.00 &  4.32 &  8.12 &  6.32 &  8.22 & \textbf{13.74} \\
      Bicyclist     & 23.27 & 33.27 & 29.28 & 30.10 & 33.43 & 35.95 & \textbf{37.24} \\
      Cart          & 48.93 & 50.97 & 57.25 & 56.07 & 50.42 & 58.40 & \textbf{60.41} \\
      Bench         & \textbf{59.72} & 54.31 & 54.46 & 56.90 & 33.94 & 50.39 & 55.14 \\
      Umbrella      & 33.83 & 43.19 & 37.70 & \textbf{47.67} & 43.83 & 45.05 & 43.53 \\
      Box           & 37.70 & 41.53 & 37.81 & 39.00 & 37.83 & \textbf{41.85} & 41.16 \\
      Pole          & 49.19 & 51.44 & 45.62 & 55.36 & 51.36 & 53.33 & \textbf{58.11} \\
      Street Lamp   & 54.91 & 62.13 & 57.19 & 59.98 & 53.63 & 59.79 & \textbf{62.43} \\
      Traffic Light & 37.27 & \textbf{43.37} & 35.53 & 39.24 & 39.10 & 42.30 & 42.38 \\
      Traffic Sign  & 38.10 & 40.84 & 40.18 & \textbf{42.58} & 39.29 & 41.22 & 39.14 \\
      Car Stop      & 35.26 & 34.95 & 29.77 & 34.53 & 32.99 & \textbf{35.88} & 33.85 \\
      Color Cone    &  4.80 & 13.87 & 18.47 & 23.83 & 15.55 & 24.42 & \textbf{25.21} \\
      Sky           & \textbf{91.10} & 87.84 & 77.93 & 87.43 & 84.55 & 88.02 & 89.68 \\
      Ground        & 87.26 & 88.51 & 87.05 & 85.62 & 89.32 & 88.19 & \textbf{89.44} \\
      Road          & 90.27 & 91.24 & 90.32 & 91.10 & \textbf{91.32}& 91.23 & 90.10 \\
      Sidewalk      & 54.07 & 58.01 & 58.43 & 57.28 & 57.58 & 57.65 & \textbf{59.63} \\
      Curb          & 44.55 & 48.44 & 46.54 & \textbf{50.88} & 48.53 & 48.41 & 50.81 \\
      Terrain       & 75.63 & 78.70 & 76.69 & 75.71 & 77.75 & 78.03 & \textbf{79.30} \\
      Vegetation    & 79.57 & 80.07 & 75.81 & 79.34 & 78.52 & \textbf{80.14} & 79.10 \\
      Building      & 76.40 & 78.45 & 75.86 & 77.18 & 76.27 & 77.68 & \textbf{79.76} \\
      \midrule
      Mean IoU (\%) & 50.67 & 53.90 & 51.38 & 54.36 & 51.59 & 54.52 & \textbf{56.42} \\
      \bottomrule\bottomrule
    \end{tabular}
  \end{adjustbox}
\end{table}

\textbf{Results on MVSeg Dataset.}
Table~\ref{tab:metrics2} further evaluates performance on the more challenging MVSeg benchmark. Under the RGB-T setting, our model achieves the best results with 72.14 Dice, 78.32 FwIoU, and 56.42\% mIoU, showing clear improvements over both traditional fusion models (e.g., RTFNet, EGFNet) and modern multi-stage networks (e.g., MVNet variants). Compared to DeepLabv3+ (RGB-T), our approach improves mIoU by +4.83\%, indicating more effective multimodal integration. Additionally, we observe that the RGB-T version consistently outperforms the RGB-only counterpart, verifying the complementary benefit of thermal imagery under diverse and adverse conditions. These results demonstrate the generalizability and robustness of our framework across different multispectral datasets. Tables \ref{tab:per-class-miou-part1-mvseg} and \ref{tab:per-class-miou-part2-mvseg} report per-class IoU on the MVSeg test set. In Part I, seven state-of-the-art RGB--T fusion methods (CCNet, OCRNet, STM, LMANet, MFNet, RTFNet, EGFNet) achieve mean IoUs between 51.63\% and 53.44\%, with EGFNet the strongest. In Part II, classic architectures (FCN, PSPNet, DeepLabv3+) and their MVNet variants reach mean IoUs of 50.67\%--54.52\%, while our MSTF-Net leads at 56.42\%, a +1.90\% gain over the previous best MVNet(DeepLabv3+). MSTF-Net delivers especially large improvements on Cart (+2.01\%), Pole (+4.78\%), Bench (+4.75\%) and Street Lamp (+2.34\%), and consistent \textgreater 1\% boosts on Sidewalk, Curb and Bicyclist. These results demonstrate that our multi-scale spatiotemporal fusion and cross-modal attention mechanisms substantially enhance representation of both fine-grained objects and global scene structures.

\begin{figure}[htbp]
  \centering

  \begin{subfigure}[t]{0.48\linewidth}
    \centering
    \includegraphics[width=\linewidth]{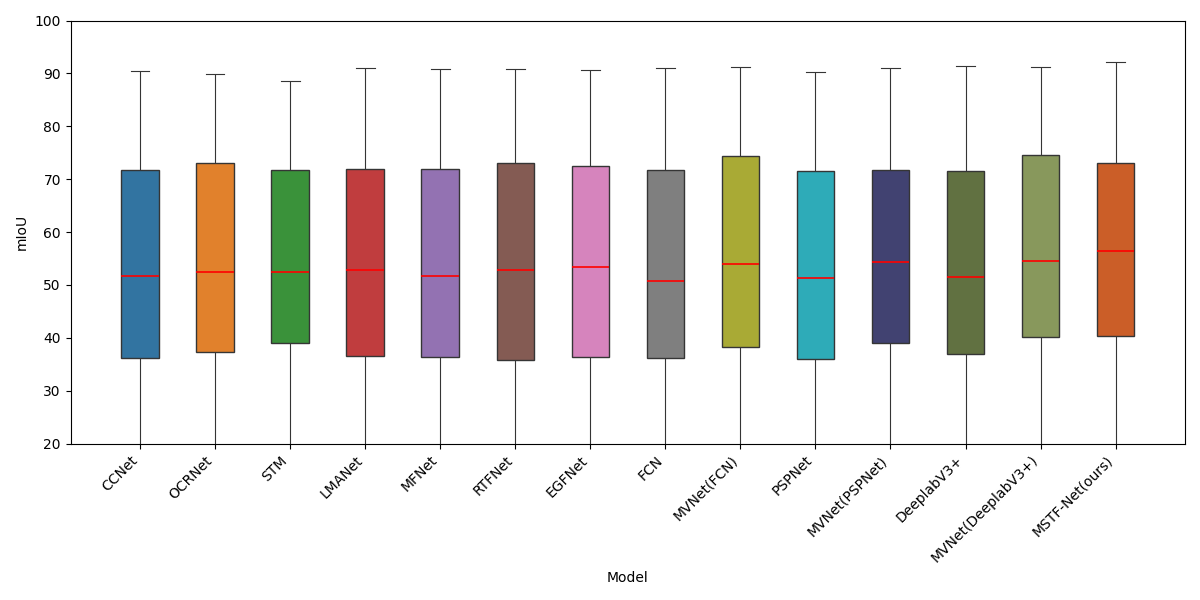}
    \caption{}
    \label{fig:mvseg_table}
  \end{subfigure}
  \hfill
  \begin{subfigure}[t]{0.48\linewidth}
    \centering
    \includegraphics[width=\linewidth]{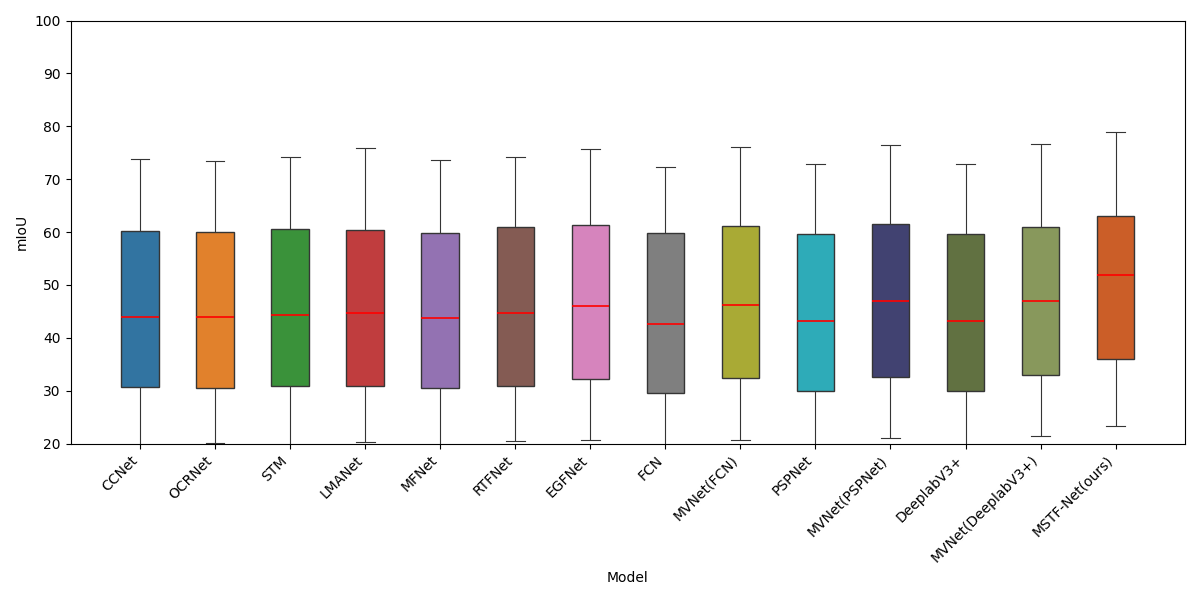}
    \caption{}
    \label{fig:cart_table}
  \end{subfigure}

  \caption{Quantitative comparison tables.
           (a) MVSeg performance;
           (b) CART performance.}
  \label{fig:qualitative_tables_vertical}
\end{figure}

Additionally, Fig.~\ref{fig:qualitative_tables_vertical} demonstrates the miou of different models

\begin{table}[ht]
  \centering
  \scriptsize
  \caption{Experimental results of the ablation study of the MTCSC. The best value of each metric is \textbf{bolded}.}
  \label{tab:MTCSC-ablation}
  \begin{adjustbox}{width=\textwidth}
  \begin{tabular}{l|cc|cc|cc}
    \toprule  \toprule
    & \multicolumn{2}{c|}{FCN} 
    & \multicolumn{2}{c|}{PSPNet} 
    & \multicolumn{2}{c}{DeeplabV3+} \\
    \cmidrule(lr){2-3}\cmidrule(lr){4-5}\cmidrule(l){6-7}
         & w/o MTCSC & with MTCSC 
                & w/o MTCSC & with MTCSC 
                & w/o MTCSC & with MTCSC \\
    \midrule
    Acc on MVSeg dataset (\%)  & 66.25 & 66.98
                               & 67.30 & 68.05
                               & 67.79 & \textbf{68.11} \\
    Acc on CART dataset (\%)   & 64.15 & 64.98
                               & 65.12 & 65.67
                               & 65.23 & \textbf{65.87} \\
    mIoU on MVSeg dataset (\%) & 52.25 & 53.02
                               & 53.32 & 54.10
                               & 55.12 & \textbf{55.37} \\
    mIoU on CART dataset (\%)  & 48.55 & 49.37
                               & 49.80 & 50.25
                               & 50.21 & \textbf{50.63} \\
    Inference time (ms)        & \textbf{16.35} & 16.75
                               & 20.10 & 20.50
                               & 16.87 & 17.23 \\
    \bottomrule \bottomrule
  \end{tabular}
  \end{adjustbox}
\end{table}
\begin{figure}[htbp]
  \centering
  \begin{subfigure}[b]{0.95\linewidth}
    \centering
    \includegraphics[width=\linewidth]{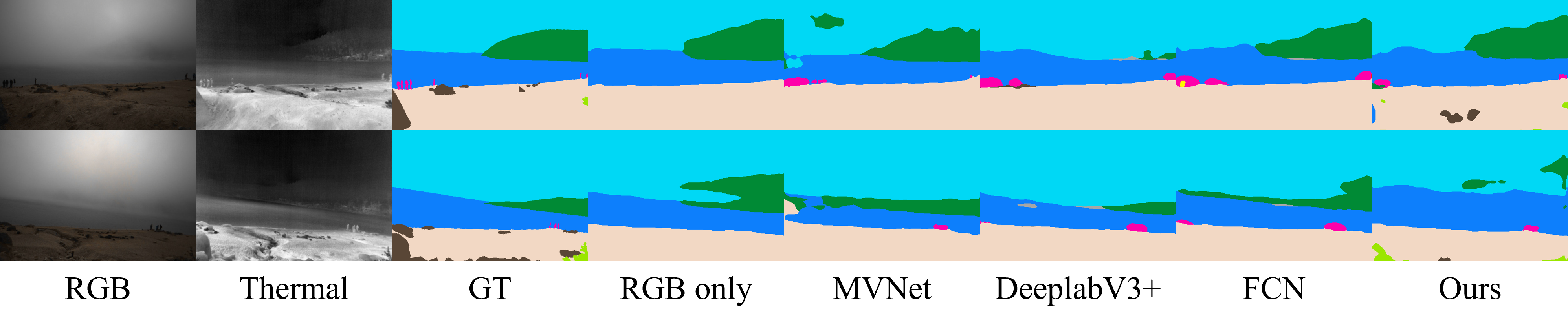}
    \caption{}  
    \label{fig:cart1}
  \end{subfigure}

  \vspace{1em}

  \begin{subfigure}[b]{0.95\linewidth}
    \centering
    \includegraphics[width=\linewidth]{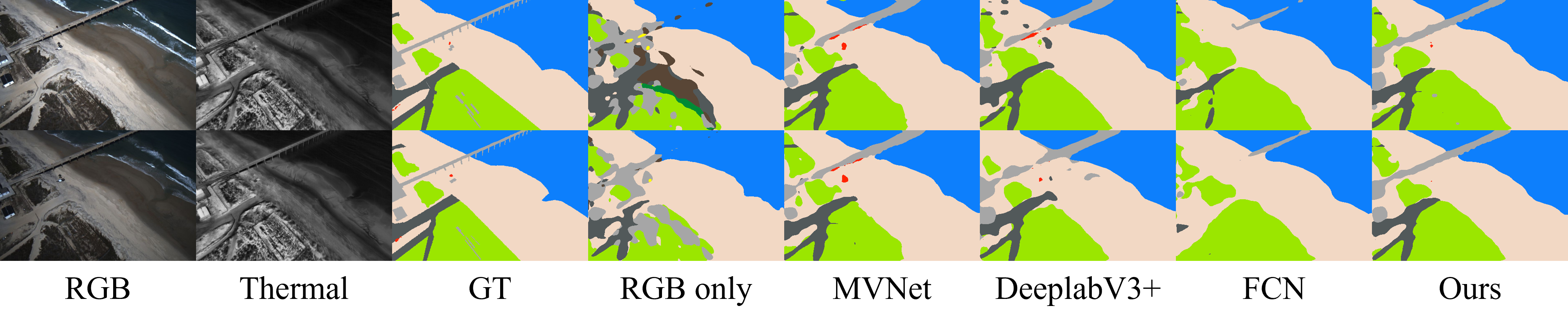}
    \caption{}  
    \label{fig:cart2}
  \end{subfigure}

  \vspace{1em}

  \begin{subfigure}[b]{0.95\linewidth}
    \centering
    \includegraphics[width=\linewidth]{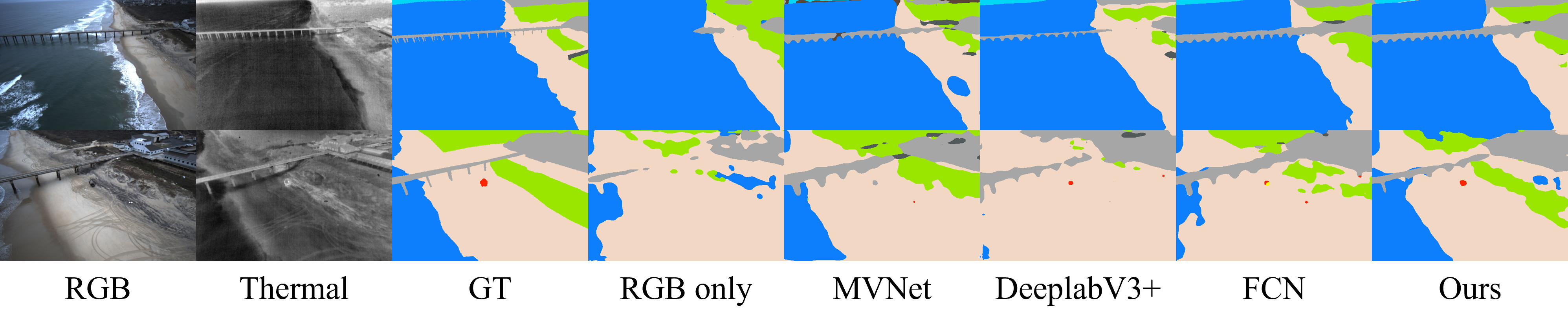}
    \caption{}  
    \label{fig:cart3}
  \end{subfigure}

  \vspace{1em}

  \begin{subfigure}[b]{0.95\linewidth}
    \centering
    \includegraphics[width=\linewidth]{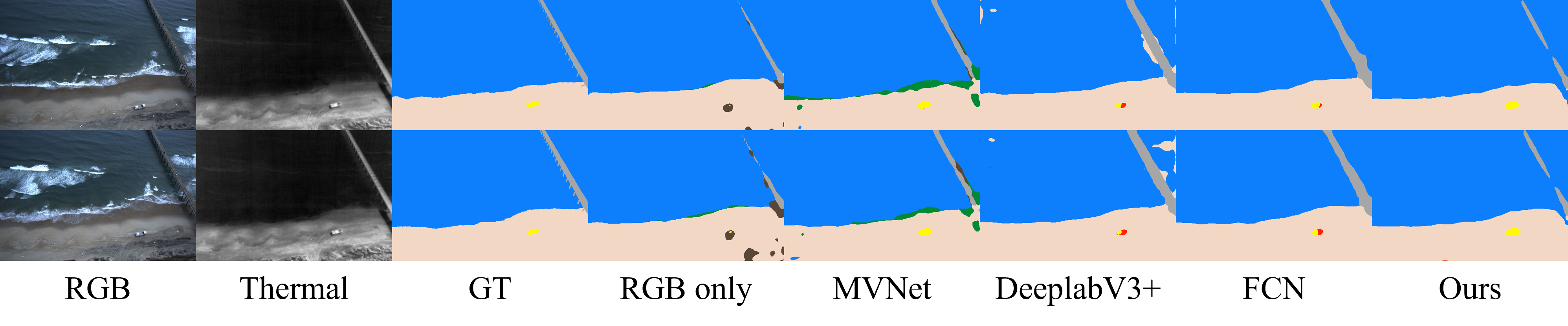}
    \caption{}  
    \label{fig:cart4}
  \end{subfigure}

  \vspace{1em}

  \begin{subfigure}[b]{0.95\linewidth}
    \centering
    \includegraphics[width=\linewidth]{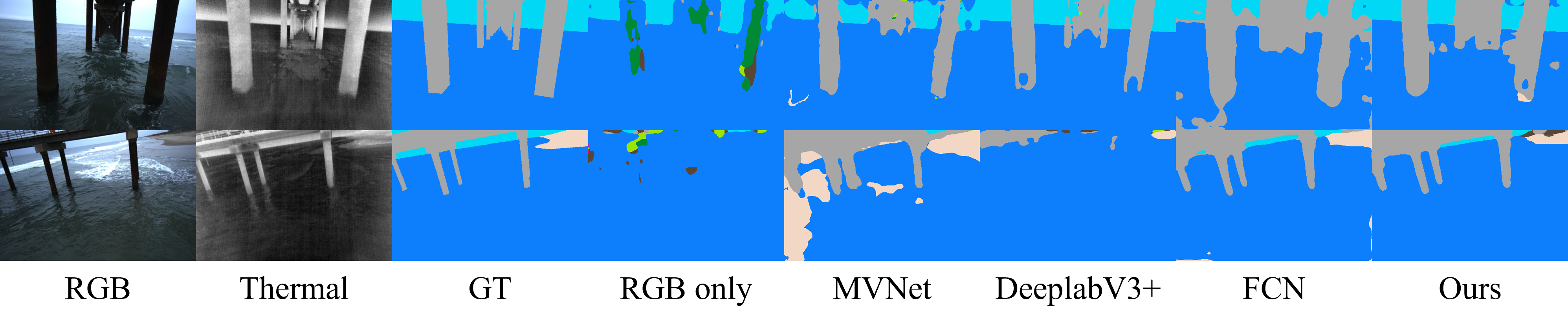}
    \caption{}  
    \label{fig:cart5}
  \end{subfigure}

  \caption{Qualitative results on the CART. (a) Picture taken in the morning. (b) Picture taken in high altitude. (c) Picture taken with altitude change. (d) Picture taken with moving object. (e) Picture taken with fast movement.   }
  \label{fig:qualitative_cart_vertical}
\end{figure}

\begin{figure}[htbp]
  \centering
  \begin{subfigure}[b]{0.95\linewidth}
    \centering
    \includegraphics[width=\linewidth]{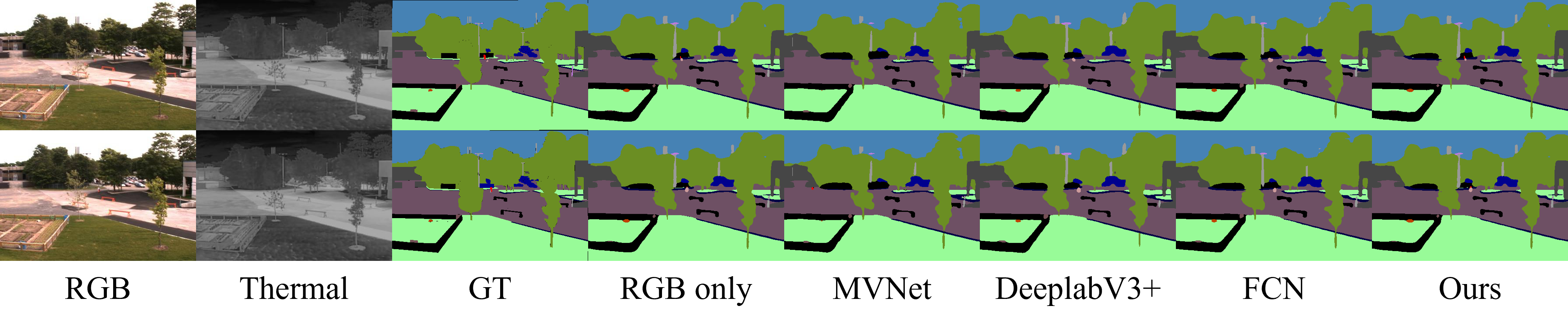}
    \caption{}  
    \label{fig:mvseg1}
  \end{subfigure}

  \vspace{1em}

  \begin{subfigure}[b]{0.95\linewidth}
    \centering
    \includegraphics[width=\linewidth]{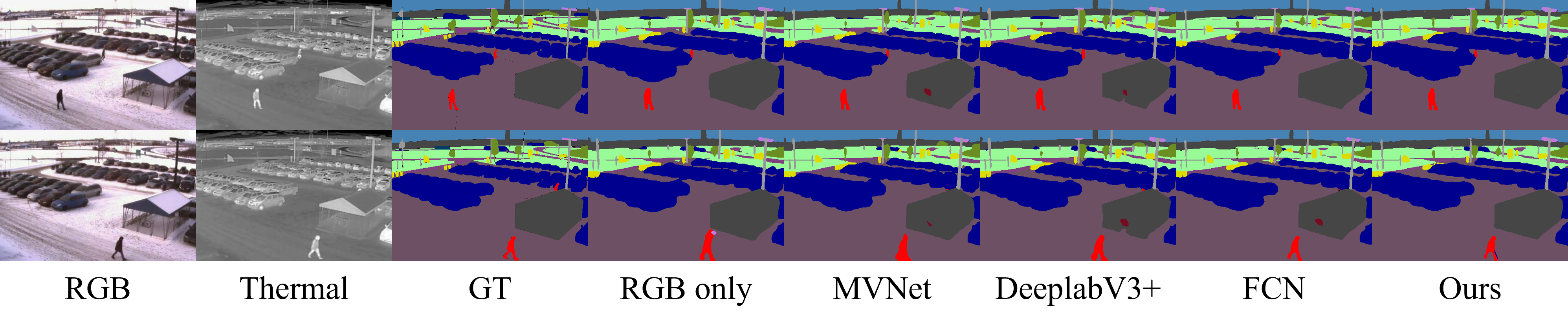}
    \caption{}  
    \label{fig:mvseg2}
  \end{subfigure}

  \vspace{1em}

  \begin{subfigure}[b]{0.95\linewidth}
    \centering
    \includegraphics[width=\linewidth]{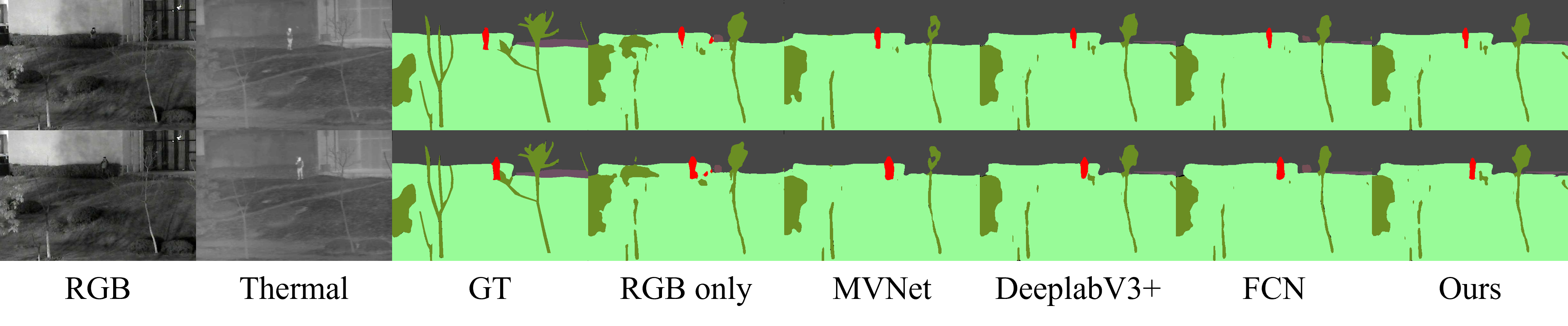}
    \caption{}  
    \label{fig:mvseg3}
  \end{subfigure}

  \vspace{1em}

  \begin{subfigure}[b]{0.95\linewidth}
    \centering
    \includegraphics[width=\linewidth]{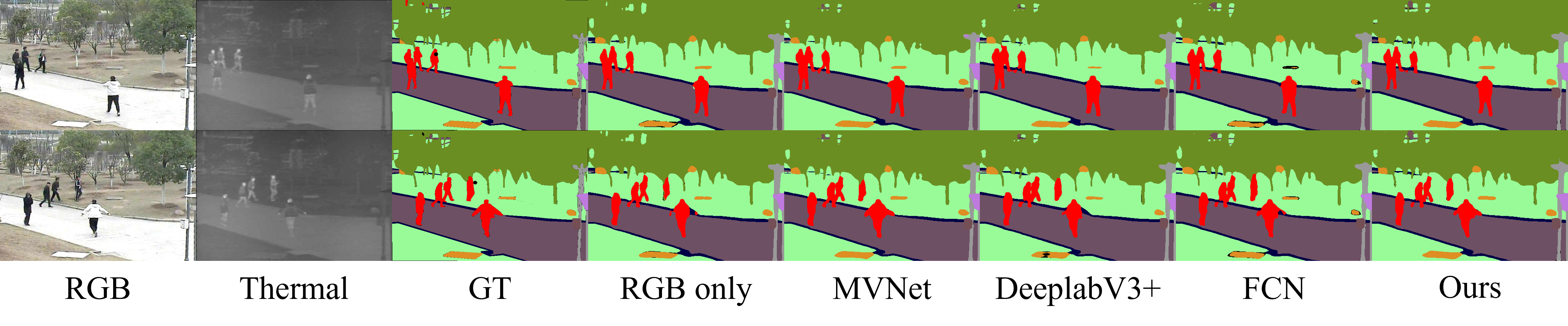}
    \caption{}  
    \label{fig:mvseg4}
  \end{subfigure}

  \vspace{1em}

  \begin{subfigure}[b]{0.95\linewidth}
    \centering
    \includegraphics[width=\linewidth]{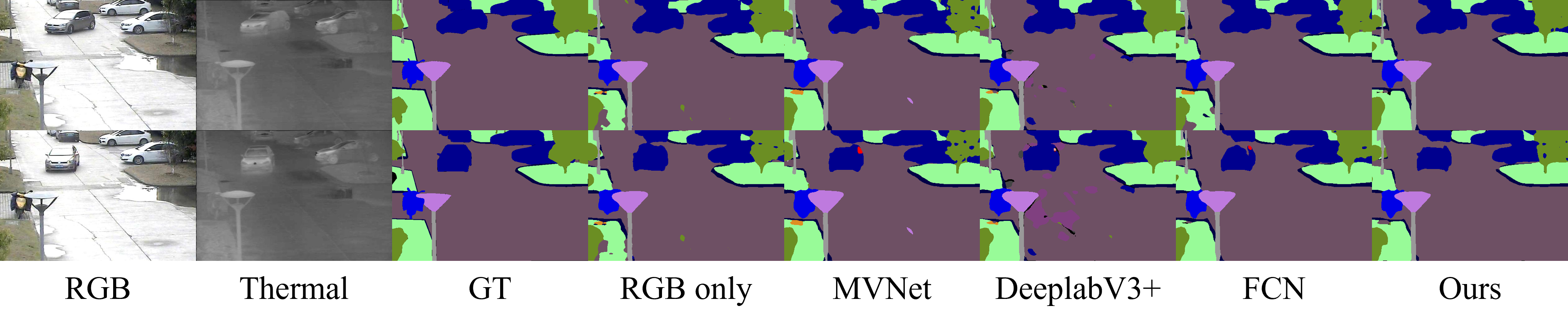}
    \caption{}  
    \label{fig:mvseg5}
  \end{subfigure}

  \caption{Qualitative results on MVSeg. (a) Picture taken in the morning. (b) Picture taken in the snow. (c) Picture taken in the night. (d) Picture taken with multiple moving objects. (e) Picture taken in the rain.}
  \label{fig:qualitative_mvseg_all}
\end{figure}

\subsection{Qualitative Results}

We present qualitative results on CART and MVSeg datasets in Fig.~\ref{fig:qualitative_cart_vertical} and \ref{fig:qualitative_mvseg_all}. Our method consistently produces more accurate and coherent segmentation maps compared to baselines. As shown, MSTF-Net effectively captures fine structures such as bridges and walkways, and preserves small objects under complex conditions.

In dynamic scenes, our predictions exhibit strong temporal consistency with minimal flickering or missing objects, benefiting from multi-scale temporal modeling. Furthermore, MSTF-Net better suppresses modality noise, maintaining clear semantic boundaries even in visually degraded inputs.
These results highlight the robustness of our approach in handling spatial ambiguity, cross-modal noise, and temporal variation.

\subsection{Ablation Study}

\begin{table}[ht]
  \centering
  \scriptsize
  \caption{Experimental results of the ablation study of the backbone. The best value of each metric is \textbf{bolded}.}
  \label{tab:backbone-ablation}
    \begin{adjustbox}{width=\textwidth}
  \begin{tabular}{l|cccc}
    \toprule \toprule
    Backbone                   & ResNet50 & ResNet101     & ResNet152      & Mini-inception \\
    \midrule
    Acc on MVSeg dataset (\%)  & 65.40    & 65.82         & \textbf{66.15} & 61.17          \\
    Acc on CART dataset (\%)   & 63.56    & 64.02         & \textbf{64.48} & 59.43          \\
    mIoU on MVSeg dataset (\%) & 51.43    & 51.92         & \textbf{52.65} & 48.08          \\
    mIoU on CART dataset (\%)  & 47.89    & 48.24         & \textbf{48.96} & 44.67          \\
    Inference time (ms)        & 15.90    & 19.74         & 25.13          & \textbf{11.57} \\
    \bottomrule \bottomrule
  \end{tabular}
  \end{adjustbox}
\end{table}

To validate the contributions of our backbone choices, the MSCSE module, and the MTCSC module, we conducted three sets of ablation experiments on the MVSeg and CART datasets. Tables~\ref{tab:backbone-ablation}--\ref{tab:MTCSC-ablation} summarize accuracy (Acc), mean IoU (mIoU), and single-frame inference time for each configuration.

\subsubsection{Backbone Ablation}

Table~\ref{tab:backbone-ablation} compares four backbones---ResNet50, ResNet101, ResNet152, and Mini-inception---on both datasets.

\begin{itemize}
  \item \textbf{Accuracy \& mIoU.}  
    As depth increases, ResNet's performance steadily improves.  
    On MVSeg, Acc rises from 65.40\,\% (ResNet50) to \textbf{66.15\,\%} (ResNet152), and mIoU from 51.43\,\% to \textbf{52.65\,\%}.  
    On CART, Acc climbs from 63.56\,\% to \textbf{64.48\,\%}, and mIoU from 47.89\,\% to \textbf{48.96\,\%}.  
    Mini-inception runs fastest (11.57 ms) but yields notably lower accuracy and mIoU than ResNet152.
  \item \textbf{Inference Time.}  
    Deeper ResNets incur higher latency: ResNet152 takes 25.13 ms per frame vs.\ 15.90 ms for ResNet50.
\end{itemize}

\textbf{Conclusion:} ResNet152 delivers the best segmentation quality, while ResNet50 or Mini-inception may be preferable for latency-sensitive applications.

\subsubsection{MSCSE Module Ablation}
\label{subsec:mscse}

\begin{table}[ht]
  \centering
  \scriptsize
  \caption{Experimental results of the ablation study of the MSCSE. The best value of each metric is \textbf{bolded}.}
  \label{tab:MSCSE-ablation}
  \begin{adjustbox}{width=\textwidth}
  \begin{tabular}{l|cc|cc|cc}
    \toprule  \toprule
    & \multicolumn{2}{c|}{FCN} 
    & \multicolumn{2}{c|}{PSPNet} 
    & \multicolumn{2}{c}{DeeplabV3+} \\
    \cmidrule(lr){2-3}\cmidrule(lr){4-5}\cmidrule(l){6-7}
         & w/o MSCSE & with MSCSE 
                & w/o MSCSE & with MSCSE 
                & w/o MSCSE & with MSCSE \\
    \midrule
        Acc on MVSeg dataset (\%)  & 64.37 & 66.25
                               & 64.82 & 67.30
                               & 65.58 & \textbf{67.79} \\
    Acc on CART dataset (\%)   & 62.56 & 64.15
                               & 62.88 & 65.12
                               & 63.12 & \textbf{65.23} \\
    mIoU on MVSeg dataset (\%) & 51.43 & 52.25
                               & 52.78 & 53.32
                               & 53.85 & \textbf{55.12} \\
    mIoU on CART dataset (\%)  & 47.89 & 48.55
                               & 49.10 & 49.80
                               & 49.94 & \textbf{50.21} \\
    Inference time (ms)        & \textbf{15.90} & 16.35
                               & 19.74 & 20.10
                               & 16.23 & 16.87 \\
    \bottomrule \bottomrule
  \end{tabular}
  \end{adjustbox}
\end{table}

Table~\ref{tab:MSCSE-ablation} shows the effect of adding MSCSE to three base networks (FCN, PSPNet, DeepLabV3+).

\begin{itemize}
  \item \textbf{Accuracy \& mIoU Gains.}
    \begin{itemize}
      \item FCN: MVSeg Acc +1.88\,\% (64.37$\rightarrow$66.25), CART Acc +1.59\,\%; mIoU +0.82--0.66\,\%.
      \item PSPNet: MVSeg Acc +2.48\,\%, CART Acc +2.24\,\%; mIoU +0.54--0.70\,\%.
      \item DeepLabV3+: MVSeg Acc \textbf{67.79\,\%}, CART Acc \textbf{65.23\,\%}; mIoU up to \textbf{55.12\,\%} (MVSeg) and \textbf{50.21\,\%} (CART).
    \end{itemize}
  \item \textbf{Inference Overhead.}  
    Adding MSCSE increases latency by only 0.3--0.6 ms across all networks.
\end{itemize}

\textbf{Conclusion:} MSCSE consistently boosts both accuracy and mIoU with negligible impact on inference speed, demonstrating its effectiveness at modeling multi-scale context.

\subsubsection{MTCSC Module Ablation}
\label{subsec:mtcsc}

Table~\ref{tab:MTCSC-ablation} reports results when integrating MTCSC on top of MSCSE-enhanced backbones.

\begin{itemize}
  \item \textbf{Additional Accuracy \& mIoU Gains.}
    \begin{itemize}
      \item FCN + MTCSC: MVSeg Acc +0.71\,\%, CART Acc +0.83\,\%; mIoU +0.77--0.82\,\%.
      \item PSPNet + MTCSC: MVSeg Acc +0.75\,\%, CART Acc +0.55\,\%; mIoU +0.78--0.45\,\%.
      \item DeepLabV3+ + MTCSC: MVSeg Acc \textbf{68.11\,\%}, CART Acc \textbf{65.87\,\%}; mIoU up to \textbf{55.37\,\%} (MVSeg) and \textbf{50.63\,\%} (CART).
    \end{itemize}
  \item \textbf{Inference Overhead.}  
    MTCSC adds only ~0.4 ms to each model's latency.
\end{itemize}

\textbf{Conclusion:} MTCSC delivers further improvements on top of MSCSE, confirming the benefit of multi-task cross-semantic coupling with minimal extra cost.

\textbf{Overall}, through a progressive ablation---backbone depth, MSCSE, then MTCSC---we achieve steady and significant gains in segmentation accuracy and mIoU on both MVSeg and CART datasets, while keeping inference times within real-time bounds.




\section{Conclusions}
In this paper, we presented MSTF-Net, a unified framework for UAV-based multi-spectral video segmentation that explicitly addresses the dual challenges of modal fusion discrepancy and temporal appearance variation. Our proposed MSCSE module suppresses modality-specific noise and enhances cross-modal complementarity through instance query generation, consistency constraints, and residual-guided filtering. Furthermore, the MTCSC module captures both short-term details and long-term semantic trends by dynamically adjusting temporal granularity based on frame distance. Extensive experiments demonstrate that MSTF-Net achieves superior segmentation performance under diverse and challenging UAV scenarios. This work provides a robust and generalizable solution for enhancing semantic perception in multi-spectral aerial video understanding.

\end{document}